\documentclass[runningheads,twocolumn]{llncs}

\usepackage{graphicx}
\usepackage{amssymb}
\usepackage{caption}
\usepackage{subcaption}
\usepackage{multicol}
\usepackage{listings}
\usepackage{algpseudocode}
\usepackage{hyperref}
\usepackage{fullpage}
\usepackage[legalpaper, margin={0.5in,1.0in}]{geometry}
\begin{document}

\twocolumn[{\centering{\Huge Three Practical Workflow Schedulers for Easy Maximum Parallelism\par}\vspace{3ex}
	{\Large David M. Rogers$^\dagger$\orcidID{0000-0002-5187-1768}\par}\vspace{2ex}
	\par\vspace{4ex}}
{\centering\bfseries Abstract\par}
\smallbreak
Runtime scheduling and workflow systems are an increasingly popular algorithmic
component in HPC because they allow full system utilization
with relaxed synchronization requirements.
There are so many special-purpose tools for task scheduling,
one might wonder why more are needed.
Use cases seen on the Summit supercomputer needed better integration with MPI and
greater flexibility in job launch configurations.
Preparation, execution, and analysis of computational chemistry
simulations at the scale of tens of thousands of processors revealed
three distinct workflow patterns.
A separate job scheduler was
implemented for each one using extremely simple and robust designs:
file-based, task-list based, and bulk-synchronous.
Comparing to existing methods shows unique benefits of this
work, including simplicity of design, suitability for HPC centers,
short startup time, and well-understood per-task overhead.
All three new tools have been shown to scale to full utilization
of Summit, and have been made publicly available
with tests and documentation.
This work presents a complete characterization of the minimum effective task granularity
for efficient scheduler usage scenarios.
These schedulers have the same bottlenecks, and hence similar
task granularities as those reported for existing tools following comparable paradigms.
\medbreak
$^\dagger$ {Oak Ridge National Laboratory, Oak Ridge TN 37831, USA} \\
\par\vspace{2ex}]

\authorrunning{D. M. Rogers}


\begin{tabular}{ll}
& {\bf Abbreviations Used} \\
API & application programming interface \\
HPC & high-performance computing \\
DAG & directed acyclic graph \\
DFM & distributed free monoid \\
REST & representational state transfer \\
METG & minimum effective task granularity
\end{tabular}\\

\section{ Introduction} 

  There can be no doubt that task scheduling is one of the core infrastructure
components of HPC.  At a system-level, an effective scheduler fills up the parallel
computer with work, and obtains statistics on resource utilization.
At a program-level, algorithms are constantly being re-written for task parallelism.
For example, job dispatching is a key component of the OpenMP runtime,
and C++ standards are expanding with parallel constructs like threads, co-routines,
and futures.\cite{HPX}

  One of the major difficulties facing the widespread adoption of task scheduling
methods is the lack of uniformity in their intended usage and user interface.
This is especially important for task systems, because task completion is a
synchronization event that we would ideally like to exploit for a variety of other
work such as logging, network and disk I/O, and, of course,
inspection of results and creation of new tasks.  This requires strong guarantees
that all the expected outputs of a task are visible before completion events
are triggered.

  From this perspective, run-time task schedulers that make specific
assumptions about task outputs can become easier to use and implement.
This work introduces three new scheduling implementations that provide
fully functional, minimal archetypes for user interactions with workflow schedulers.
The major distinctions between them come from alternative assumptions
about where and how task outputs are specified.
The {\tt pmake} tool is file-based,
and uses a single managing process to push jobs to workers.
The global view of jobs allows an earliest-finish-time priority.
It synchronizes based on presence or absence of output files.
The {\tt dwork} tool is network-service based.  Worker processes
interact with a task-list server to retrieve and record completion of tasks
by name.  Its view of the task graph lends itself to a FIFO scheduling strategy.
It synchronizes through a single server that guarantees
all dependencies of a task are complete before serving that task to any worker.
The {\tt mpi-list} tool maintains a unique assignment of data elements to
processes, so that no synchronization is needed for local operations.
It targets single-program, multiple-data operations within the MPI
(message-passing) paradigm of synchronization.

  These concepts have all appeared before in multiple forms.  What is new
and novel here is that they have been stripped to their essentials and made
as simple as possible.  Limiting each tool
to implement only a single synchronization mechanism makes
the assumptions clear and simplifies the user interface, debugging process,
and system installation.
  
  Task scheduling systems are productivity tools when they are easy
to use, and so fitness-for-purpose is most important.  However, they can
also be sources of overhead for a large computation.
I quantify the overhead of each task scheduling tool
in order to understand the `minimum effective task granularity' (METG).
This measure was introduced by Ref~\cite{taskbench}
to measure the overhead incurred during actual task processing.
Basically, it measures (in seconds) the task difficulty needed to equally divide total execution
time between scheduling overhead and actual work done on the task.
Any task taking a bit longer than this will spend the majority of its
time in the computation phase.
They found that there are large number of scheduling systems that
achieve METG values of 0.01 to 0.1 milliseconds (ms).  However,
general-purpose task schedulers like Spark, Dask, and Swift/T
require hundreds of ms, likely due to maintaining additional job metadata.

  The results section of this work will show that the underlying assumptions
of the task distribution method can change the key parameters in the scaling
equation.  Specifically, the METG can depend on the machine size.
The design of the task manager determines the functional form of that dependence.
At the configurations tested here (1-6912 ranks), the METG varied
between 0.5 ms and 5 seconds depending on the tool used.  In practice,
this means that certain types of schedulers require larger batches of work
to be practical.

The tools presented here have already proven useful on several projects
underway on the Summit supercomputer, including managing hundreds of molecular dynamics
simulations and analysis steps {\tt pmake},\cite{covid} running docking and AI-based
rescoring ({\tt dwork}),\cite{docking} and summarizing results ({\tt mpi-list}).\cite{dataset}
This work shows that, when used correctly, they achieve full parallelism
with negligible run-time overhead.

Section~\ref{s:review} provides a brief classification of
some popular runtime task schedulers, showing key distinctions
in their implementation ideas.  Section~\ref{s:design} presents
the design of three new task schedulers, {\tt pmake}, {\tt dwork},
and {\tt mpi-list}.  Section~\ref{s:eval} presents the details
of this work's evaluation method to quantify METG.  Section~\ref{s:results}
provides numerical results, which are discussed further in sections~\ref{s:discuss}
and~\ref{s:concl}.

\subsection{ Review of existing schedulers}\label{s:review}

  There are multiple trade-offs implicit in the design and selection
of schedulers.  Ref.~\cite{rsilv16} provides a taxonomy
and Ref.\cite{rsilv16} gives an extensive feature comparison.
Table~\ref{t:features} compares features of schedulers
considered for the applications motivating this work.\cite{covid}
Note that the list is heavily biased toward applications
intended to run datacenters because I am interested in tracking progress
of `campaigns' in a fault-tolerant way.  For this work, I define
a simulation campaign as a collection of many compute-intensive
tasks.  In some cases, the entire campaign can collectively require more resources than
available from the batch scheduler (e.g. due to limits on available nodes and run-times).
Task managers can achieve fault tolerance over campaigns
by tracking the list of pending tasks and tasks resulting in errors.
The first three columns in the table show that most workflow managers
built for campaigns also provide database backends and an API to query task status.
This fault tolerance comes at the cost of increased latency for tracking and
assigning tasks.

  Setting up and interacting with the task manager is also important.
The `Language' column shows what programming languages are available to
create task graphs.
Most task systems are focused tasks forming a
directed acyclic graph (DAG), and interface with Python.
Most scientific computing programs, however, use a mix of C, C++, Fortran, and Python,
and not all computations can be expressed as a DAG.
The `Dynamic' column shows how the task manager implements
updates to the task graph from the computation itself (while in-progress).
Most task managers support appending tasks to the graph,
but few consider this as a design feature.
Some have a loop construct to explicitly support iterating a task multiple times.
Fireworks allows general rewrites of the task graph, but does
not provide a consistent execution semantics.
Regent provides constructs to dynamically utilize
hierarchical processor groups.

  The last column describes the way tasks are assigned to workers.
In pull-based systems, the workers request tasks from the manager.
These systems are easier to setup than push-based managers, since
the manager process does not need a node-list at startup.
Push-based systems like Spark and Dask attempt to optimize
task placement. Signac and RADICAL-Pilot use a push system so that groups of nodes
can be allocated at once.

\begin{table*}
{\centering
\begin{tabular}{*7{l}}
   & Target & Query & Persistence & Language & Dynamic & Push/Pull \\
   \hline
Apache Airflow\cite{airflow} & datactr & REST/CLI & SQL & Py & modify globals\cite{airflowdyn} & pull from broker \\
Apache Spark\cite{spark}  & datactr & CLI & checkpoint & Scala/Py/R/Java & interactive & push \\
{\tt mpi-list}$^*$ & datactr & no & no & Py & interactive & push \\
Resque\cite{resque}  & datactr & REST & Redis & Ruby & append & pull \\
RADICAL-Pilot\cite{radical} & modeling & ? & Mongo & Py & append & push \\
MetaQ\cite{metaq} & modeling & no & file & shell & no & push \\
Signac/flow\cite{signac} & modeling & CLI/WebUI & file or Mongo & Py & append & push \\
SnakeMake\cite{snakemake} & modeling & CLI & file & shell/Py & no & push \\
{\tt pmake}$^*$ & modeling & CLI & file & shell & no & push \\
Fireworks\cite{fireworks}  & modeling & CLI/WebUI & Mongo & Py & rewrite & pull \\
Pegasus\cite{pegasus}   & modeling & CLI/WebUI & SQL & DSL/Py & ? & push \\
{\tt dwork}$^*$  & modeling & ZMQ/CLI & TKRZW & Proto+ZMQ & replace & pull \\
Dask.distributed\cite{dask}  & analytics & Py API & no & Py & ? & push/steal \\
Regent\cite{regent}  & HPC & no & no & Regent & hierarchical & shared \\
Parsec\cite{parsec}  & HPC & no & no & DSL & loop & shared \\
\hline
\end{tabular}
\caption{Feature comparison of present work ($^*$) with popular workflow schedulers.
Missing entries (?) indicate features that were not able to be determined from the tool's documentation.
datactr: cloud or HPC datacenters, REST: HTTP interface, CLI: command-line interface,
WebUI: web user interface, Py: Python, DSL: domain-specific language}\label{t:features}
}
\end{table*}

In operational terms, the three task schedulers introduced in this work,
({\tt pmake}, {\tt dwork}, and {\tt mpi-list}) are closest to SnakeMake,
RADICAL-Pilot, and py-sparkling\cite{pysparkling}, respectively.  Key differentiating features
of the present tools include a smaller API and better integration with the
MPI-inside-batch job paradigm appropriate for Summit and Andes systems.
{\tt pmake} uses all resources within a single batch job, avoiding indefinite
batch queue waiting times and maximizing multi-node usage.
{\tt dwork} has a single client and server.  This relies on the user to launch
clients with the appropriate resources instead of specifying them to the workflow system.
Neither py-sparkling nor Spark distribute datasets over MPI ranks.

  Apache Airflow, {\tt dwork}, Signac/flow, SnakeMake, Fireworks, Pegasus, and Dask.distributed
are all built around the basic idea of providing a central database of tasks
with task-dependencies forming a DAG.  Most provide a programmatic way
of constructing tasks and specifying their dependent tasks.  Python and yaml
fit naturally into this level of abstraction because of Python's design goal as an
easy shell interface and its large ecosystem of libraries.  Most tasks at this level
of abstraction are expected to last around tens of seconds.
{\tt dwork} also implements the DAG idea.  It allows for dynamic tasks
by implementing a `rewrite' mechanism to add new dependency edges
to a running task.  This replaces the running task back into the queue to be re-run when
the newly added dependencies are complete.

  Resque, RADICAL-Pilot, and MetaQ do not maintain graph dependencies for tasks.
Instead, tasks are expected to be trivially parallel.  Resque and RADICAL-Pilot do provide
the option of submitting other, follow-on, tasks.
Resque tasks interface well with Ruby, since they can be yaml-serialized
Ruby function code.
Its intended use case is to carry out long-running tasks needed
to manage files, eventually updating github's web databases.
RADICAL-Pilot is intended to run Python and shell tasks.
MetaQ provides a Slurm-like syntax for specifying resources.
It is intended to bundle multiple batch-queue jobs into a single batch job
to reduce overall queue wait times and increase (outer) batch job sizes.

  Spark and {\tt mpi-list} are unique in that they do not use the DAG-of-tasks concept.
Instead, work distribution is based on a globally known assignment
of data to processors where computations are carried out.  Because of this,
its programming model is bulk-synchronous single-program, multiple data.
Their method of task-addition is described as `interactive', since the program
dynamically chooses its execution path at the top-level of execution.

  Regent and Parsec target HPC workloads.
These are not expected to out-live a job allocation,
so do not have database backends.  They also
employ special distribution strategies to
optimize for minimal latency between tasks.
Specifically, both Regent and Parsec use a shared execution model
where every worker explicitly models (at least) the
local task graph and its mapping to compute resources.
Leaving out a central database reduces communication latencies.

  Compared to the existing runtime task scheduling systems,
{\tt pmake} fits into the class of file-directed, make-like job schedulers.
This design decision is a natural parallel to the Make build tool.
The {\tt mpi-list} tool follows the general ideas of Spark, but
is more compatible with HPC environments.
In particular, it is based on mpi4py\cite{mpi4py}, simplifying job
launch.  Instead of a resilient distributed dataset holding a list
of partitions, {\tt mpi-list} uses a DFM,
which holds a list of arbitrary objects.  This makes it operate
equally well holding lists of plain integers, numpy or cupy arrays or pandas DataFrames.
The DAG-of-tasks concept underlying {\tt dwork}
is not unique.  Instead, what {\tt dwork} provides is a less obtrusive
management layer.  The only purpose of {\tt dwork} is to
define a network API for creating and assigning tasks.
This borrows a design idea from {\tt resque}, and provides
a much simpler framework for feature expansion than other
available options.

\section{ Description of the Design and User Interface}\label{s:design}

  This section describes the design and implementation
of three separate task schedulers.  The designs differ
based on their intended use pattern.

\subsection{ pmake}\label{s:pmake}

  Pmake is a parallel version of the `Makefile' concept.
Every task corresponds to one or more output files, which
determine whether the task needs to be run.
Rules describe how to create output files from input files.
Pmake is a single Python program that reads a list of rules
and a list of target files and runs all tasks in the task-graph.

Because the pmake process views the entire task graph, it
is able to assign earliest start times to all tasks by traversing the
DAG from leaf (immediately executable) to root (most dependent)
nodes.  Instead of using the time directly, it uses the total
node-hours consumed by a task and all its transitive
successors to assign a priority to every task.
Then, it uses a greedy strategy to choose the highest
priority task from those runnable at each time point.

\begin{figure*}
\begin{subfigure}{1.0\textwidth}
\begin{lstlisting}[language=Python]
simulate:
  resources: {time: 120, nrs: 10, cpu: 42, gpu: 6}
  inp:
    param: "{n}.param"
  out:
    trj: "{n}.trj"
  setup: module load cuda
  script: |
    {mpirun} simulate {inp[param]} out{[trj]}

analyze:
  resources: {time: 10, nrs: 1, cpu: 1}
  inp:
    trj: "{n}.trj"
  out:
    npy: "an_{n}.npy"
  setup: module load Python/3
  script: |
    {mpirun} Python compute_averages.py {inp[trj]} {out[npy]}
\end{lstlisting}
\caption{Example {\tt rules.yaml} file with make-rules for executing parallel programs.}\label{p:rules}
\end{subfigure}

\begin{subfigure}{1.0\textwidth}
\begin{lstlisting}[language=Python]
sim1:
  dirname: System1
  out:
    npy: "an_0.npy"
  loop:
    n: "range(1,11)"
    tgt:
      npy: "an_{n}.npy"
\end{lstlisting}
\caption{Example {\tt targets.yaml} file listing out high-level output files.}\label{f:targets}
\end{subfigure}
\caption{Pmake program inputs for a typical simulate, then analyze workflow.
This workflow creates {\tt System1/an\_0.npy}, $\ldots$, {\tt System1/an\_10.npy}
by running the simulate and analyze rules multiple times
from the {\tt System1} directory.}\label{f:pmake}
\end{figure*}

In place of the Makefile, {\tt pmake} employs a {\tt rules.yaml} file.
An example {\tt rules.yaml} is shown in listing~\ref{p:rules}.
The two rules form a sequence {\tt simulate} $\to$ {\tt analyze},
which eventually produces files like {an\_1.npy}, {an\_2.npy}, and so on.
In operation, {\tt pmake} will generate {\tt simulate.$n$.sh} from the
setup and script section and execute it in the background
sending its stderr and stdout to {\tt simulate.$n$.log}.
It continues until it runs out of available allocated compute nodes.
Exiting scripts release their nodes.
Scripts exiting with a zero-return value trigger any waiting rules.

There are several important design decisions shown from this example.
Rules have extra meta-data compared to standard makefiles:
{\em i}) a resource set, {\em ii}) a list of multiple
input and output files, {\em iii}) a setup script, and
{\em iv}) automatic creation of an \{mpirun\} command,
which expands to the appropriate {\tt srun} or {\tt jsrun},
depending on whether Slurm or LSF scheduler is used.
Also, the syntax for variable substitution is determined by Python's
{\tt format()} function.
For rules that can make multiple output files,
one variable is allowed, and is defined by matching on names in the
{\tt out} section.

A resource set specifies a division of the allocated
nodes for a job into equally-sized resources -- each with
a fixed number of CPUs and GPUs.  Usually, one MPI rank
is assigned to each resource set, but it is also possible
to set {\tt ranks = R} to launch $R$ MPI ranks
on each resource set instead.  Resources in {\tt pmake}
also include time (specified in minutes).
This is used by {\tt pmake} to prioritize tasks ready
to be run on the machine by estimating earliest-finish time.

The {\tt inp} section lists files required before
the rule can be triggered.  Like {\tt make}, {\tt pmake}
stops searching for rules when it finds all the files
needed to build its outputs.  Inputs can also be specified
using a {\tt loop} directive (not shown), which lists input files
generated by filling in a template with a Python iterable.

Often the same set of operations are run across
multiple problem instances.  Listing~\ref{f:targets}
shows a {\tt targets.yaml} file.  This file lists out
the top-level targets the user would like to build.
Each target has an arbitrary name and attributes available
to be substituted into rules that run for the target.
Reserved keywords for the target are {\tt dirname},
{\tt out}, and {\tt loop}.  All the target files
are relative to the {\tt dirname}.  The {\tt out}
and {\tt loop} file lists have the same format as for inputs
to rules.

When a rule is run, its setup and job-scripts are concatenated
together and pre-pended by a {\tt set -e} and a {\tt cd}
into the target's {\tt dirname}.  The result is written
to a shell script `rulename.n.sh' named after the rule name and its
template variable, $n$, (if present).  That shell script
is executed locally by a call to {\tt popen},
and its output stored in a logfile, also named `rulename.n.log'
after the rule.

The use of Python's format function allows Python code
to be spliced into the script.  Substitution happens in
order from targets to rules, so that variable references
will only work for variables declared earlier. The order is:
{\em i}) members of the target (other than loop -- these are first, so not substituted),
{\em ii}) variables in the loop directive are substituted sequentially,
{\em iii}) members of the rule (other than script),
{\em iv}) script directive (which also gets `{mpirun}` defined from the scheduler).
One drawback is that braces (\{\}) must be escaped.

\subsection{ dwork}\label{s:dwork}

  {\tt dwork} is a client/server API implementing a bag of tasks.
Because tasks can have dependencies, these tasks can form a
directed acyclic graph (DAG).  When tasks insert new tasks,
the computation expressed by the graph can create loops.
Table~\ref{t:api} lists the key messages implemented by the
{\tt dwork} API.  For the implementation, each of these messages
is encoded in Google protocol buffers\cite{protobuf} and
passed through ZeroMQ.\cite{zeromq}

Because the server is not assumed to have a complete view of
the task graph, it uses a first-in-first-out assignment strategy
for tasks.  Workers that request a task are served with the oldest task
inserted into the database that is ready to run
(based on completed dependencies).  On the other hand, tasks
that are re-inserted back into the graph are added to the front of the
priority queue.  This double-ended queue setup is exactly the same one
used for work-stealing, where remote workers resume the `oldest' task,
while the local processor executes the `newest'.

The server for {\tt dwork} ({\tt dhub}) uses a TKRZW\cite{tkrzw} database to store the task graph.
Like Redis\cite{redis} it can save and restore the database to file for persistent state.
Unlike Redis, its API is centered around creating tasks and assignments.
The task database internally contains only two tables:
a table of join counters and successors for each task
and a table of task metadata (name, originator, etc.) for each task.
Other run-time information, such as the list of tasks ready to run, can be
generated from these tables on startup.

I also provide a command-line tool ({\tt dquery}) as an example client
that can interact with the API from shell scripts.
Usually, users write their own software to interact with {\tt dhub}.
This works well in practice, since tasks are software anyway,
and protobuf and ZeroMQ are supported in a very wide variety
of programming languages.

\begin{table*}
{\centering
\begin{tabular}{llcp{2in}}
Query               & Parameter & Response & Description \\
\hline
Create              & Task, [Task]  &  -      & Create a new task with the given dependencies. \\
Steal                & Worker         & Task? $\vert$ Exit & Deque (steal) a ready task to be run by the worker. \\
Complete          & Worker, Task & -  & Notify the scheduler that a task is complete. \\
Transfer            & Worker, [Task] &  -      & Replace the task and add new dependencies. \\
Exit                  & Worker         &  -      & Notify the queuing system that a processing device is down.
\end{tabular}
\caption{Minimal API for maintaining a distributed a task list.
Workers send these queries to the task manager, which replies with the response.
Repetitions of zero or more entries (lists) are denoted by brackets.
The data-type for Worker is implemented as a string,
while Tasks are defined as protocol buffer messages to allow passing additional
meta-data about the task.}\label{t:api}
}
\end{table*}

The API is consumed both by users of the system (who create tasks),
and processing nodes (who call Steal, Complete and Transfer).
Figure~\ref{f:kwork} provides pseudocode for the operation
of the client and server around these API calls.
Internally, the scheduler maintains the successors of every task,
along with a join-counter.  The join counter goes to zero when
all the task's dependencies are marked completed.
It also maintains a mapping from Worker-s to sets of Task-s assigned
to that worker.
The scheduler will not assign a task (via responding to Steal)
unless all of the task's dependencies have been marked completed.

  If workers have unique hostnames, calls to
`Exit' can be run by the worker or by the user
to recover from a node failure or abort.
When receiving such a notification, the queuing system moves
tasks assigned to the exited worker back into the pool of ready tasks.

  This implementation signals completion of the task graph by
responding three different ways to `Steal'.  Usually, a task is provided
to the worker.  In case no tasks are ready, the manager responds with
a `NotFound' message.  In case all tasks are complete, the manager
responds with `Exit.'

  Valid task graphs do not contain cycles.
This is mostly guaranteed by the syntax for creating tasks.
The only potential way to add a cycle is during `Transfer'.
The transfer operation moves a task back from a worker
(where it was assigned) to the manager.
New dependencies (pre-requisites) can be added
to the task at this point.  If the new dependencies
are waiting for a task that transitively depends on
the transferred Task itself, this is a user-error
that creates deadlock.  Observationally, such tasks
will never enter the `ready' state, and thus never be served
to workers.

\begin{figure*}
\begin{multicols}{2}
\begin{algorithmic} 
 \Function{client-loop}{}
 \While{server responds with task}
        \State copy-in task inputs
        \State execute task
        \If{task error?}
          \If{self-diagnostic fails}
             \State inform server of Exit
             \State exit
          \Else
             \State inform server of error
          \EndIf
        \Else
          \State inform server of completion
        \EndIf
  \EndWhile
\EndFunction
 \Function{Steal}{}
    \If{ready tasks?}
      \State pop a task from the ready-list
      \State mark as assigned to client
      \State send to client
    \ElsIf{waiting tasks?}
       \State send NotFound
    \Else
       \State send Exit
    \EndIf
 \EndFunction
 \Function{Complete}{Worker, Task}
    \State delete assignment of task to worker
    \If{success?}
        \State mark successors ready
    \Else
        \State add successors recursively to errors set
    \EndIf
 \EndFunction
 \Function{Exit}{Worker}
   \State move node's assigned tasks back to ready set
 \EndFunction
 \Function{Create}{Task, Dependencies}
   \If{unfinished dependencies?}
     \State add new task to successor list of all dependencies
     \State add to waiting list
    \Else
      \State add to ready list
    \EndIf
 \EndFunction
 \Function{Transfer}{Task, Dependencies}
   \State delete assignment of task to worker
   \If{unfinished dependencies?}
     \State add new task to successor list of all dependencies
     \State add to waiting list
    \Else
      \State add to ready list
    \EndIf
 \EndFunction
\end{algorithmic}
\end{multicols}
\caption{Client/Server Interaction API Implementation in {\tt dwork}. %
Production client code would use an assembly-line pattern to overlap these 4 steps.}\label{f:kwork}
\end{figure*}

\subsection{mpi-list}\label{s:mpilist}

The {\tt mpi-list} tool is a Python package that provides a
functional API for manipulating lists.  The syntax and design
is inspired by Spark,\cite{spark} although the startup and communication
mechanisms of {\tt mpi-list} are purely based on mpi4py.\cite{mpi4py}
Because all MPI ranks execute the same operations on their local
portion of the dataset, its scheduling model is bulk-synchronous-parallel.

Mpi-list provides only two classes - a `Context' to hold the
MPI communicator information, and a `DFM' object to
represent distributed lists.  DFM stands for distributed free monoid.
The `DFM' object stores only the set of list elements local to each rank.
The global list is logically maintained in an ordered state,
with a contiguous and ascending subset of the list
assigned to each rank.
New `DFM' objects are created with `Context.iterates(N)',
which creates a distributed list of $N$ sequential integers.
Rank $p$ of $P$ stores the subsequence starting at $p \; \mathrm{int}(N/P) + \mathrm{min}(p, N\;\mathrm{mod}\;P)$.

  It is not usually necessary to index local list elements directly.
Instead, operations like `DFM.map(f)'.  Create new lists by
applying the function, $f$, to each list element.  There
are also functions for both full reduction and parallel prefix-scan
reduction.  To move elements between ranks, a `DFM.repartition'
and `DFM.group' are implemented.

\begin{figure*}
\begin{lstlisting}[language=Python]
from mpi_list import Context
import xarray as xr

C = Context()
dfm = C . iterates(N) \
            . flatMap( lambda n: read_scored(data, prot, n) ) \
            . map( best_scores )
n = dfm.len()
if C.rank == 0:
    print(f"Read {n} pq files to {C.procs} processes in {t1-t0} secs.")
ret = dfm . map(stat) . collect()
if C.rank == 0:
    print(f"Collected stats to rank 0 in {t3-t2} secs.")
    df = pd.concat(ret)
    df.to_parquet(out / "summary.pq")
    lo = df.loc['min']
    hi = df.loc['max']

# broadcast histogram parameters
lo,hi = C.comm.bcast((lo,hi), root=0)
H = Hist(lo, hi, 301, 201)
ret = dfm . map( lambda df: H.his2d(df, 'score', 'r3') ) \
          . reduce( npsum, 0 )
if C.rank == 0:
    print(f"Collected histogram1 in {t3-t2} secs.")
    ret.to_netcdf(str(out / "score_rf3.nc"))
\end{lstlisting}
\caption{Production code snippet showing use of {\tt mpi-list} to read
a dataset of parquet files and create a 2D histogram in parallel.
This example read 2592 parquet files (80 gigabytes compressed) to 320 processes
(20 nodes of OLCF Andes) in 4.0 seconds, collected stats in 1.5 seconds,
then output its histogram in 0.4 seconds.}\label{f:mpilist}
\end{figure*}

  The `repartition' function does not simply move list elements,
but instead treats each list element as a complex list-like datastructure
containing multiple records.  This is because the intended usage of
{\tt mpi-list} is to store {\tt numpy} or {\tt cupy} arrays, or {\tt pandas} dataframes.
The repartition function thus requires three functions: one reporting
the length of the stored objects, one able to subdivide the object
into multiple chunks, and another able to combine multiple chunks together
again.  Obviously, the {\tt mpi-list} implementation is handling the
mundane counting and communication tasks.

  Similarly, the `group' function requires, as input, a function
to turn a stored object into a dictionary mapping destination list indices
to lists of objects that should be sent to that index.  {\tt mpi-list} moves
all the data to its newly determined MPI rank, and calls the user's combination
function on each new index to re-form the list back into the final output object.

  Mpi-list is released in the pypi package index, and contains full documentation
and functionality tests.

\section{ Evaluation Method}\label{s:eval}

  Tools for managing launching and logging of tasks can be measured
for performance efficiency by quantifying the overhead with
respect to sequentially running all tasks directly on a single compute resource.
Sometimes, this overhead can vary when computational resources are
actually doing work - and so it can be helpful to measure the overhead
during actual task processing.
The `minimum effective task granularity' (METG) has been introduced
as a descriptive measure in this case.\cite{taskbench}
Basically, it measures (in units of seconds) the task difficulty
needed to equally divide observed run-time
between scheduling overhead and actual work done on the task.
If the average execution time per task equals the METG, then
the total run-time (execution plus overhead)
will be twice the number of tasks times the METG.

  In practice, the effect of overhead is negligible when tasks are a few times
larger than the METG.  The METG gives a helpful guideline when dividing up work,
since in many cases task sizes can be arranged to be larger than the METG.
Also, the typical task size provides an estimate of the computational
idle time caused at the completion of a sequence of tasks.
All the task managers here operate inside a Slurm or LSF job allocation.
Hence, out of the resources allocated to the job, all but one task's worth
could remain idle for this time.

  Although the task schedulers presented here were first put to production use
on computational chemistry research with mixed CPU/GPU workloads,\cite{covid,docking,dataset}
the non-uniformity of those tasks prevent quantitative comparison.
This work presents a standard, synthetic benchmark to highlight overheads.
I chose GPU-intensive matrix multiplication because of
its importance to applications in high-performance computing.
There, large (often machine-size) matrices are divided into tiles to distribute the data
and computational work among all available resources.
I apply the three schedulers here to compute a series of
$A^T B$ operations, where $A$ and $B$ are single-precision floating point
matrices with sizes between $256$ and $8192$.
This operation appears very often as a building block of linear algebra
calculations like computation of the wavefunction
overlap matrix, $S = \psi^\dagger \psi$, in quantum mechanics.\cite{dftfe,GW}
In those applications, $\psi$ may consist of hundreds of thousands of
tiles like $A$ and $B$, giving rise to millions of tasks.
In these workflows, however, neglect the hypothetical communication steps
needed to sum the results in computing $S$.

  Each task management system was benchmarked using
a weak-scaling methodology -- where number of tasks
scaled with processors.  The scale was set to 1024 total
kernel executions per rank.  Every run used 1 MPI rank per GPU,
except for baseline runs intended to determine the minimum
compute time needed to run the kernel itself (which used only 1 GPU).
For {\tt pmake} and {\tt dwork}, tasks consisted of 256
iterations of the matrix-multiplication kernel.
For {\tt mpi-list}, one single list containing all problems was
created, and then the kernel was run inside a map-function.
In practice, this allowed each rank to run its 1024 assigned
kernel runs inside a for-loop.

  Note that this division means that all workflow systems
encounter the startup costs of spawning MPI jobs,
initializing the GPU and allocating memory only once
-- except for {\tt pmake}.
{\tt pmake} encounters the batch spawn and
`alloc' cost four times per rank.  These costs
cannot reliably be overlapped with computation,
and determine {\tt pmake}'s METG.

  {\tt mpi-list}, on the other hand, has a METG
determined entirely by imbalance in execution times.
Since every process runs exactly 1024 kernels,
processes that get interrupted or have slower than usual
GPU accesses will hold back progress on all the other ranks.
Its METG is thus determined by the point at which
idle time (slowest minus fastest completion) equals
the ideal task completion time.  In other words, the
METG for {\tt mpi-list} is the ``slowest minus fastest'' completion time
per task.

  For {\tt dwork}, the overhead per-task comes from
receiving its next task via communication with the task database.
Thus, its METG is the latency time for accessing the database
{\em multiplied by} the number of MPI ranks.  The latter
multiplication happens because the database must serve
every rank with a task at a rate faster than the time needed
for one worker to complete a single task in order to keep that
worker busy.

Table~\ref{t:versions} lists the software and dependency library versions
for the scheduling tools tested in this work.  The calls to cublas-sgemm
were made via blaspp for {\tt pmake} and {\tt dwork}, or via
{\tt cupy} for {\tt mpi-list}.

\begin{table}
\begin{subfigure}{0.3\textwidth}
{\centering
\begin{tabular}{ll}
Package & Version \\
\hline
pmake & commit 05f727 \\
\hline
blaspp\cite{blaspp} & 2021.04.01 \\
cuda & 11.2.0
\end{tabular}
\caption{Software used by pmake and its kernel.}
}
\end{subfigure}
\hfill
\begin{subfigure}{0.3\textwidth}
{\centering
\begin{tabular}{ll}
Package & Version \\
\hline
dwork & commit 88ebb1 \\
\hline
zeromq & 4.3.3 \\
cppzmq & 4.7.1 \\
protobuf-cpp & 3.14.0 \\
TKRZW & 0.9.3 \\
blaspp\cite{blaspp} & 2021.04.01 \\
cuda & 11.2.0
\end{tabular}
\caption{Software used dwork and its kernel.}
}
\end{subfigure}
\hfill
\begin{subfigure}{0.3\textwidth}
{\centering
\begin{tabular}{ll}
Package & Version \\
\hline
mpi-list & 0.3 \\
\hline
Python & 3.7.10 \\
mpi4py & 3.0.3 \\
cupy & 8.5.0 \\
cuda & 10.1.243
\end{tabular}
\caption{Software used by mpi-list and its kernel.}
}
\end{subfigure}
\caption{}\label{t:versions}
\end{table}

Tests were conducted on the Summit supercomputer.
Each summit node has two sockets,
each socket has 3 NVIDIA(R) V100 GPUs
and 21 usable IBM power9 processor cores.
Software versions are listed in Table~\ref{t:versions}.

\section{ Scaling Results}\label{s:results}

\begin{figure*}
\centering
\includegraphics[width=0.6\textwidth]{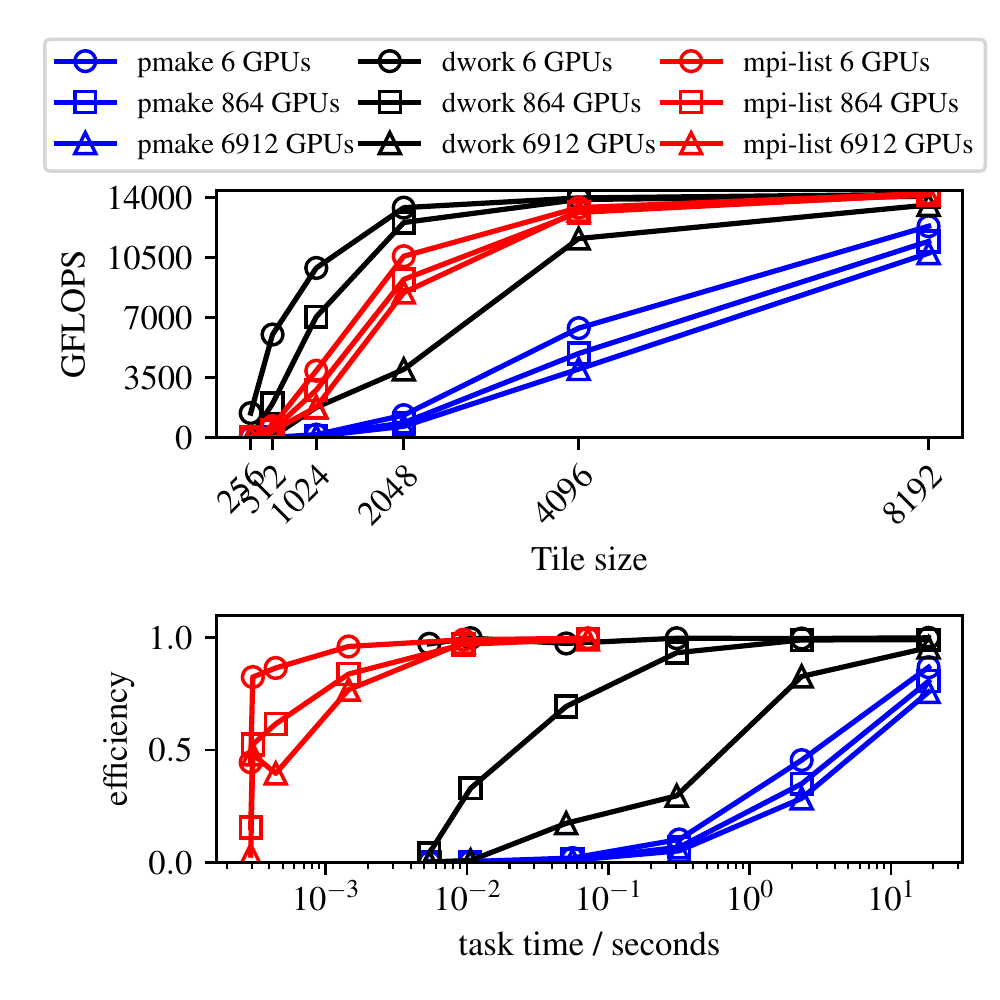}
\caption{Absolute (upper) and relative (lower) computational efficiency
per GPU measured using total time the workflow scheduler
spends in its compute phase.  One-time workflow startup phases for
each are not present in this figure, but are listed Table~\ref{t:overhead}.}\label{f:gflops}
\end{figure*}

  Fig.~\ref{f:gflops} shows computational efficiency (GFlops) of the
task systems as a function of the block size.
All task systems achieve peak theoretical efficiency for the hardware
(14 teraflops per GPU in single-precision).  This should always be
achieved when the problem sizes are large enough
so that computation time is much larger than any other overhead time
per calculation.
However, the maximum efficiency is not achieved for small tile-sizes.
This is partially because of the GPU and call path to the blas library itself,
and partially because of the overhead of the schedulers.

  Figure~\ref{f:gflops}(upper) hides potential sources of overhead
that create a difference between the single-GPU run time
and the parallel, task-scheduled run time.
Thus, all further plots are of efficiency relative to single-GPU compute time,
as is done in the lower part of Fig.~\ref{f:gflops}.
This measure highlights losses due to task scheduling,
and ignores losses due to tile-sizes that don't saturate the GPU
or (for Python) overcome the function call overhead.
The measure serves to normalize the execution time across frameworks because
there are small differences in initialization and multiplication time
(notably {\tt mpi\_list} calls the {\tt cupy} library).
The resulting plots highlight scheduler overhead instead of these details.

  The lower portion of Figure~\ref{f:gflops} reproduces the minimum effective
task granularity (METG) plot of Ref.~\cite{taskbench}.
Computational efficiency, plotted on the vertical axis,
is defined as ideal divided by actual
per-task execution time.
The horizontal axis is the task's ideal, single GPU time.
The METG is defined as the task size where computation time equals half the total
execution time per task.
It is visible as the sharp increase in efficiency
at a particular task size.
This task size is helpful to state in
terms of the ideal, `single-GPU' execution time, as we
have done here.
Note that, for {\tt pmake} and {\tt dwork}, one task is
defined as 256 iterations of the multiplication kernel.

\begin{figure}
\begin{subfigure}{0.45\textwidth}
\includegraphics[width=\linewidth]{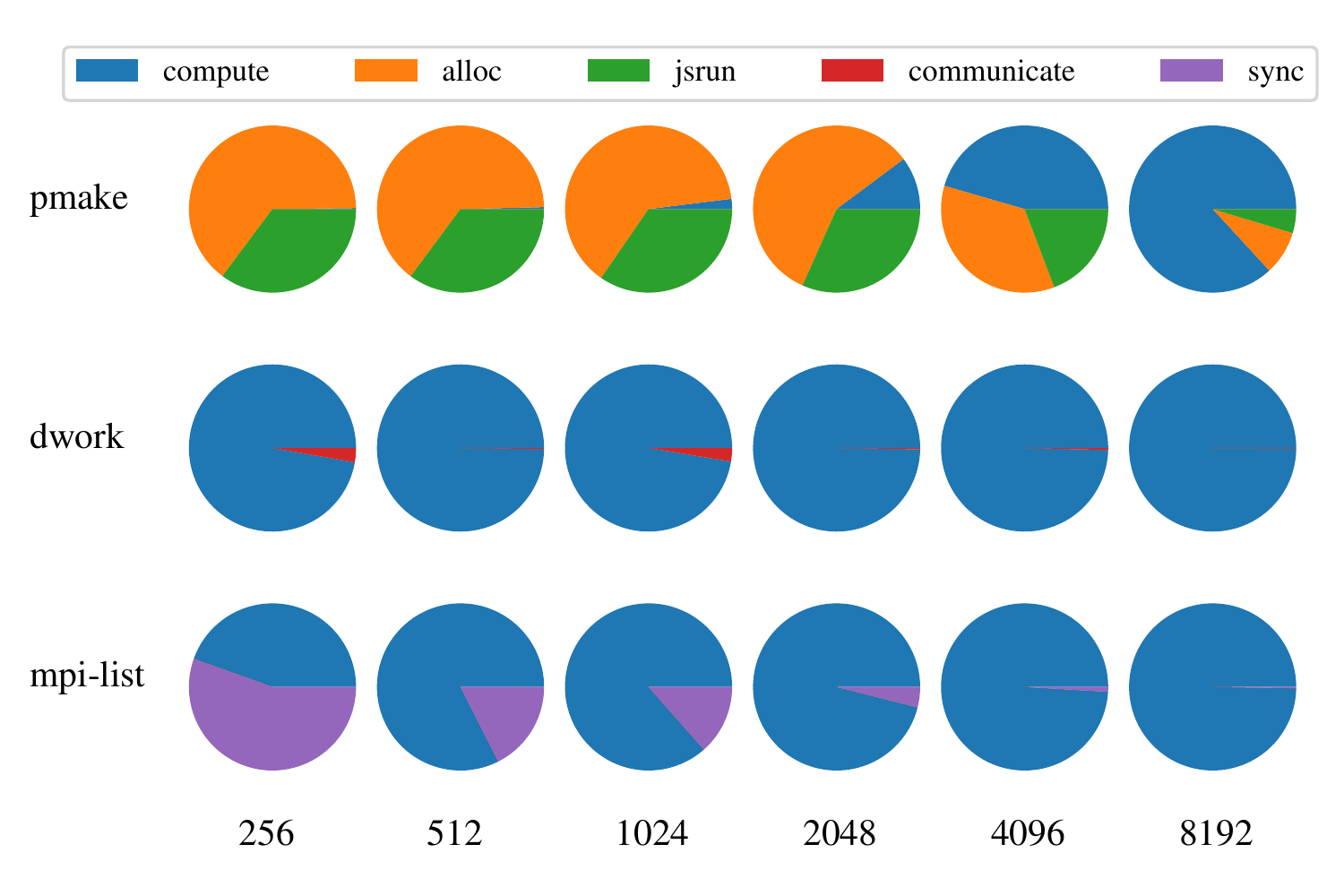}
\caption{Six ranks (single node) calculation.}\label{f:pie1}
\end{subfigure}
\hfill
\begin{subfigure}{0.45\textwidth}
\includegraphics[width=\linewidth]{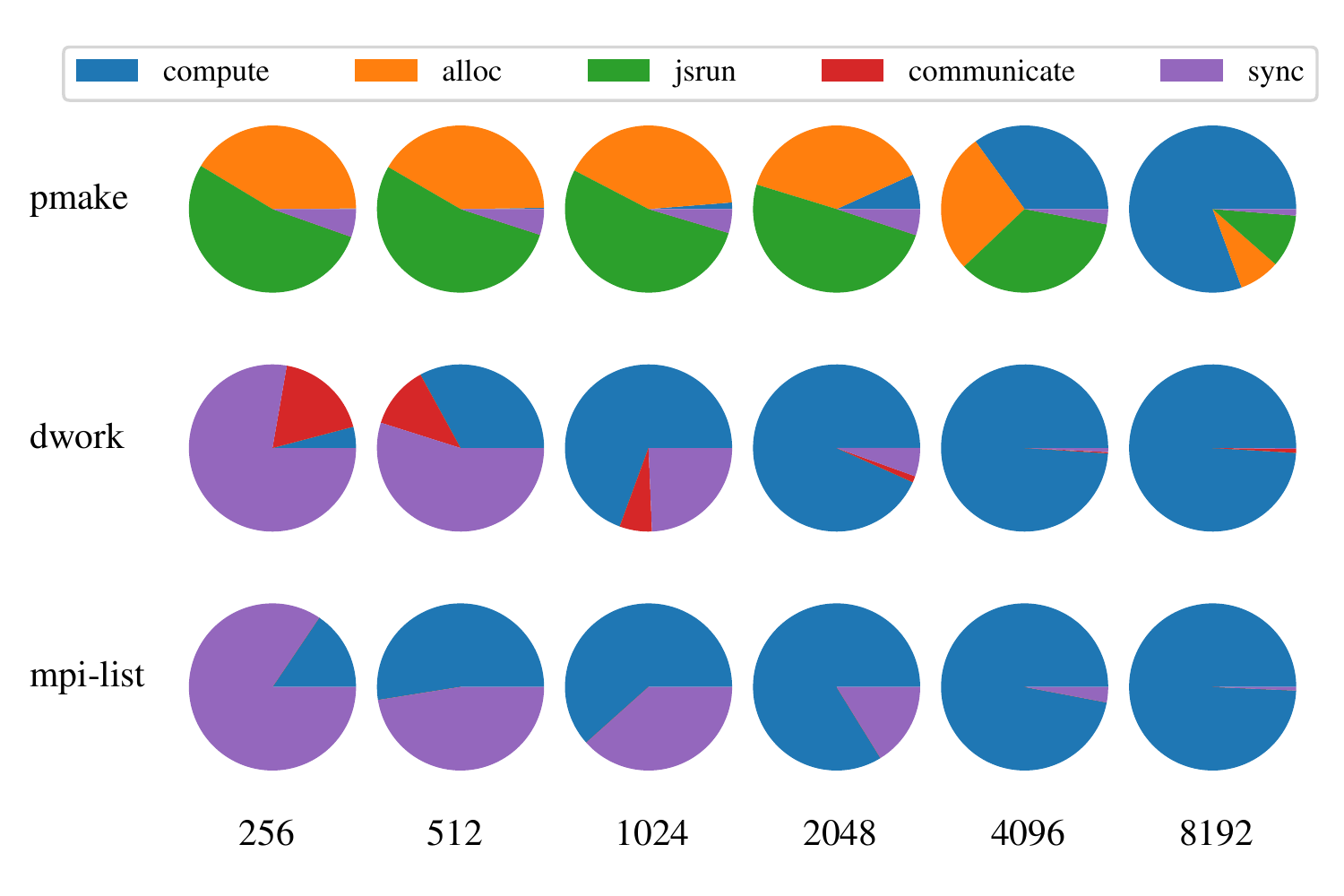}
\caption{864 ranks (144 node) calculation.}\label{f:pie144}
\end{subfigure}
\begin{subfigure}{0.45\textwidth}
\includegraphics[width=\linewidth]{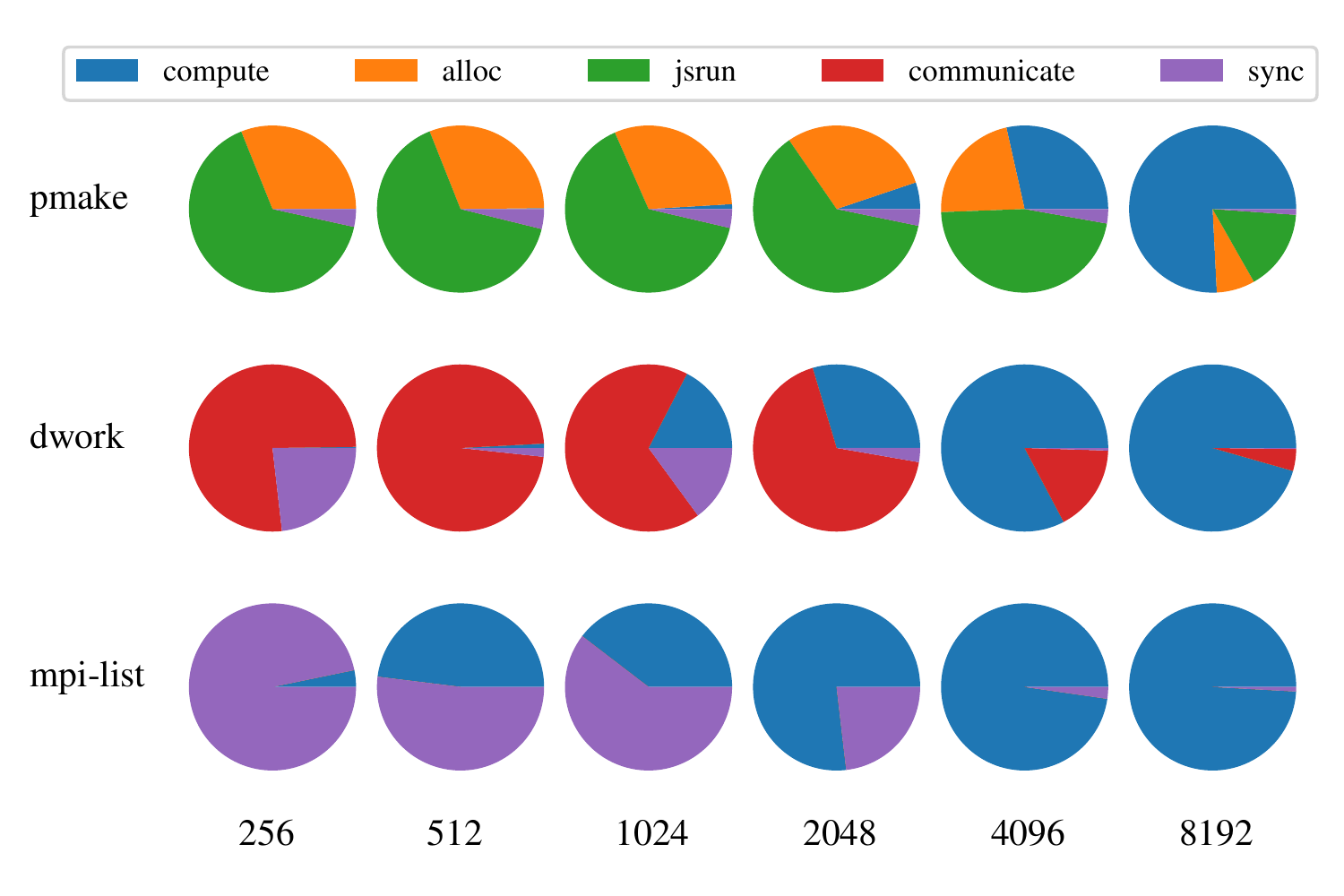}
\caption{6912 ranks (1152 node) calculation.}\label{f:pie1152}
\end{subfigure}
\caption{Pie-charts showing time breakdown between computation
and each overhead cost for each scheduling tool (y-axis).
Increasing matrix tile-sizes (x-axis)
provide enough computation to overwhelm tasking overheads.
METG can be seen as the point where the computation occupies
more than half the time.  Note that for {\tt pmake} and {\tt dwork},
each task carries out 256 matrix multiplications, so
only four tasks are sent to each rank during a run.
{\tt pmake} shows sync-time for large runs
because each pmake-task occupies 864 ranks.}\label{f:pie}
\end{figure}

Figure~\ref{f:pie} shows the breakdown of time spent per task
when running the test workflow.
Each of the three runtime scheduling tools have different
sources of per-task overhead.  The {\tt pmake} tool launches
job steps during a running allocation -- and so incurs both
overhead from the job step launch (jsrun), and from startup tasks
of the job itself (alloc).  Neither of these tasks can be overlapped
with computation easily.  The {\tt dwork} tool only stops
work to request a new task from the server, and to record completion (communication).
These two can be effectively overlapped with computation.
The {\tt mpi-list} kernel launches occur locally, so the
only overhead is at synchronization points in the code.
The `sync' time measures this end-of-job synchronization.
It is equivalent to the time difference between the fastest possible
and slowest encountered run-times per rank.

\begin{table*}\centering
\begin{tabular}{r|*{7}{r}}
ranks & {\bf jsrun time} & {\bf alloc} & \parbox{0.6in}{{\bf comm}\\ (per task)}
& \parbox{0.8in}{{\bf sync time}\\ per 1024 tasks}
& Python alloc & Python imports & dwork connection \\
\hline
   6 & 0.987 & 1.81 & 23$\mu$ & 0.09 & 2.23 & 1.05 & 1.54 \\
  60 & 1.783 & 1.81 & 23$\mu$ & 0.17 & 2.23 & 0.55 & - \\
 864 & 2.336 & 1.81 & 23$\mu$ & 0.33 & 2.23 & 2.82 & 2.74 \\
6912 & 3.823 & 1.81 & 23$\mu$ & 0.47 & 2.23 & 26.65 & 13.32
\end{tabular}
\caption{All data are times in seconds.  The symbol $\mu$ denotes a factor of $10^{-6}$.
Column headers in bold font are costs incurred per-task (depending
on the workflow tool).  Constant values were averaged over all test
runs because their timings did have significant run-to-run variation.
Other columns are useful measures of tasking system startup time.}\label{t:overhead}
\end{table*}

  Table~\ref{t:overhead} summarizes scaling measurements
of these overhead components.  Job step launch times
increase roughly logarithmically with the number of MPI ranks.
Task startup (GPU memory allocation) is constant independent
of the job size or tile size.  Communication time only appears
in {\tt dwork} when computation time is too small to hide it.
For {\tt mpi-list}, the METG equals the disparity between
fast and slow computation times.  This is shown by the
synchronization latency column.
It is slowly increasing with number of ranks.
Synchronization latency is indirectly measured for {\tt mpi-list}
by subtracting the completion time from the 1-GPU program
completion time.

  Based on the performance at 846 ranks, the METG for
  {\tt mpi-list}, {\tt dwork} and {\tt pmake} are 
0.3, 25, and 4500, milliseconds, respectively.
These can be compared to the results in Fig. 9
of Ref.~\cite{taskbench}.  There, 864 ranks correspond to between
16 and 32 nodes.  Although the interconnect and processor hardware
are different, the schedulers in that work fall into two classes
-- those with METG between 0.01 and 10 milliseconds (Chapel, Charm++, MPI, OpenMP task, OpmSS, PaRSEC, Regent, StarPU, TensorFlow, X10), and those with more than 1000 (Spark, Swift/T and Dask).
Interestingly, this work's Spark-like {\tt mpi-list} has the best performance,
probably due to the lack of file I/O.  {\tt dwork} falls in-between the two groups,
and {\tt pmake} is in the second group.

  The bulk of {\tt pmake}'s waiting time is attributable to the startup
time for executing the `jsrun` command, which allocates a group of
processors to execute an MPI program in parallel.
We can see that this increase in the METG is due to an approximately
logarithmically increasing task-job launch times.

For {\tt dwork}, the METG is negligible
on a single MPI rank, but increases
proportional to the number of ranks.
The relevant number from Table~\ref{t:overhead} is
the 23 microseconds needed for the Steal/Complete API calls.

Unfortunately, its METG scales linearly with the number of concurrent compute ranks.
For {\tt dwork}, the maximum communication value is
achieved by a kernel that does no work.  In that case, the server
is the bottleneck, and the time equals the total number of tasks assigned
times the round-trip time of the network API per per-task.
I have used a 2-level forwarding tree, where each rack of 18
Summit nodes communicates with a rack-leader.
The rack leaders forwards all messages to a single task server running on
the job's launch node.  As the number of ranks increases, the communication
time measured is this two-hop time.

\section{ Discussion}\label{s:discuss}

  The {\tt pmake} tool provides a makefile-like syntax to run a series
of {\tt srun} or {\tt jsrun} launches, as if they had been typed
into a job-script directly.  It's scaling behavior is a direct consequence
of this choice.  Because job launch can take tens of seconds,
tasks must take at least this long in order to make effective
use of the machine with this approach.  This is usually
the case when bundling together many jobs that would
otherwise have been run on their own as isolated batch scripts.
Scanning file presence and constructing a task graph
can cause delays too, but are generally worthwhile because they
can avoid duplicating work.

  The {\tt dwork} tool is built around a client/server model.
Its task dispatch latency shows typical behavior following from this model.
Message transfer rates using ZeroMQ\cite{zeromq} and hash-table entry read/write
rates form lower bounds on the latency.  I have avoided additional costs
deriving from establishing TCP connections by establishing
a tree-shaped message forwarding chain.  I have added additional
costs by wrapping transmitted tasks and messages into Google protocol
buffers.\cite{protobuf}  All of these choices combine to a measured cost of
about 23 microseconds latency per task.

Although 23 microseconds of latency seems very small,
this number means that only 44,000 tasks can be distributed
to all workers per second.  For a job with 44,000 MPI ranks, every task
must then last at least one second in order to achieve full utilization.
Even then, receiving a response could take half a second.
This waiting time can be hidden by overlapping computation and communication,
which I have implemented in the client.
With overlapping, the task granularity is controlled by the task manager's
maximum rate of dispatching jobs.

  Two simple strategies are worth pursuing for increasing task
dispatch rates further.  The first is sending multiple tasks per `Steal' request.
I have already implemented this as a separate `Steal $n$' request.
The second is to replicate the server by creating multiple task databases
that distribute tasks among themselves.  I have designed the API
to adapt to this pattern in the future, since delegating a task to another
task database is logically the same as assigning it to a worker.

  The last new task manager, {\tt mpi-list}, has characteristics
that follow from the Spark paradigm.
Execution happens where data elements sit, and
each process maintains its own list of local data elements.
It is not suited to general task graphs.
Instead, it works best on map-reduce type operations
over datasets that can fit in the machine's memory.
{\tt mpi-list} has the smallest latency per task because our
tasks do not include file operations.
In my implementation, job-launch and network
communication is based on MPI, making it easy
to use on HPC systems.  I observed
a moderately large startup time for the initial launch of Python.
This is not a per-task cost, but it is troubling that
it increases with MPI ranks.  Since this
is likely due to I/O overhead from all ranks importing libraries
on startup, future work should investigate using
the spindle tool to reduce this cost.\cite{spindle}

All of the task latencies observed for workflow
scheduling systems presented in this work are on par
with similar measurements in the literature.
{\tt mpi-list} has a latency of 0.3 ms, entirely due to
barrier synchronization costs.  This is slightly above the fastest
schedulers from Ref.~\cite{taskbench} because the synchronization
is global, and not point-to-point.
The METG for {\tt mpi-list} is the largest, around 1-5 seconds.
Even still, this is comparable to the smallest METG observed
in Ref.~\cite{taskbench} for the Spark, Dask, or Swift/T workflow systems.
{\tt dwork} has an overhead of 1 second at the largest node counts,
comparable or smaller to that measured in Ref.~\cite{taskbench}
for Regent, StarPU, and Parsec.

  Although the METG sets a strong-scaling limit,
there are many important features for workflow scheduling systems
including flexible user interface, reproducibility and tracking,
ease of installation, and simplicity.\cite{roadmap}
These multiple trade-offs necessitate making
situation-specific and qualitative comparisons.

\section{ Conclusions}\label{s:concl}

  This work has successfully implemented and used in production
several task managers for distributing asynchronous, parallel
work.  All three are very well suited for HPC centers with batch
systems and MPI.  Arguably, one of their greatest advantages
is the simplicity of the implementation and user interface.
The conditions under which each of the systems
will execute a task are straightforward.  Task tracking
is not complicated, and the minimum effective task
granularity is well-characterized.

Each of these scheduling systems serves as a base of
useful functionality which can be modified for more efficient
and specialized uses in the future.  The {\tt pmake} scheduler
is very similar to the more full-featured SnakeMake\cite{snakemake},
so {\tt pmake}'s idea of specifying a parallel resource set
and machine-dependent substitions for \{{\tt mpirun}\}
can be adopted into that code.  The {\tt mpi-list} code
seems to fill a new niche for Spark-like computations inside MPI.
Nevertheless, it can be limited by memory bounds on large datasets.
Further work can eliminate this issue without substantial changes
to {\tt mpi-list itself} by loading wrappers for the data elements
that maintain the data on disk except during processing.

The {\tt dwork} client/server model can be substantially
expanded to include features like: 1. separate pools of work
with independent servers (trivial), 2. forwarding of messages
to maintain constant open connections per rank
(implemented in the code released with this article),
3. more comprehensive display and interactivity with
the task queue (moderate),
4. shared responsibility for handing out tasks, sharded
between multiple servers (moderate),
5. advertisement of shared-memory access methods
for intermediate results (advanced).
Most of these features (when needed) would reduce
overall task latency times and further increase scalability.

  Interestingly, the minimum effective task granularity of
all three dynamic scheduling tools follows different scaling laws.
The {\tt pmake} tool launches tasks in the background,
but every task incurs startup costs from the scheduler's
resource allocation and from the program startup itself.
Its METG is thus equal to the job startup costs. 
{\tt dwork} pushes job information to running clients,
so the startup costs become negligible over long runs.
Instead, its METG comes from the overhead of communicating
the `next ready' task.  That overhead scales with the number of
concurrent workers as long as a single-task-server design is used.
So, its METG is the per-task latency times the number of workers.
The {\tt mpi-list} system apparently has no waiting time for
the next instruction.  However, it does have synchronization points
and uses statically assigned work items.  Thus, its METG
is equal to the difference in run-times between `fastest' and `slowest'
concurrent worker.  That difference depends on the number of
data elements processed, and is the subject of the study of extreme
value distributions.\cite{gumbel}

  When presented with these interfaces, task management appears easy.
Difficulties in most implementations come from additional design
requirements.  For example, building a domain-specific interface
requires associating additional information with each task.
Preventing duplication of tasks can require external
methods to determine the status of completed tasks.
Specializing tasks to run on specific hardware or at coordinated
times also requires some external delegation or
synchronization mechanism.

  This work's contributions uncouple the task schedulers from
all of these concerns.  Doing so achieves a simpler and more
flexible design.  For some problems, however, it
leaves practical implementation issues unresolved.
In particular, the work does not addressed the data lifecycle management issue.
In OpenMP, Quark, StarPU, Parsec, and their predecessors,
tasks are associated with in- and out- memory pointers.
This requires tracking the association from data locations to tasks.
Distributed memory runtime systems often add the further
step of copying data to local resources before execution can start.
Solutions like distributed caching filesystems and
tuple-spaces have been designed to address this issue.\cite{tuplespace}

  Each of the issues above have multiple potential
solutions with different trade-offs.  This strategy allows the user to
make their own implementation choices.  It reduces the complexity
of the task manager and usually also the user code.
I have shown that this approach has latency overhead
on the order of 23 milliseconds when under heavy load
and can create and deque one million task in about a minute
-- competitive with task schedulers designed for HPC workloads.

  This work does not provide a single solution to synchronization
and coordination mechanisms.  Instead, we are left with two opposing ideas.
Task graphs are functional, expressing every part of the data flow
from task to task, and tracking progress by pulling data from a
collective memory.
Bulk-synchronous operations on data
are imperative, prescribing precisely what each processor
should do.  These track progress by a global instruction counter.
One middle-way that has been left unexplored
is to earmark certain tasks for compatible locations
and separately implement a rendezvous mechanism to
synchronize the start-time of those tasks.

  Overall, these tools provide a usable, versatile way to
manage scientific simulation campaigns outside the batch scheduler.
This brings scaling benefits, bringing down the time and effort
needed to run multiple inter-related calculations.
Their unique features are simplicity of design, suitability for HPC centers,
very short startup time, and well-understood per-task overhead.

\section*{Availability}

The tools described in this work have been made available under open-source licenses
at the following locations:
\begin{itemize}
\item {\tt pmake}: \href{https://code.ornl.gov/99R/pmake}{https://code.ornl.gov/99R/pmake} (GPL v3)
\item {\tt dwork}: \href{https://github.com/frobnitzem/dwork}{https://github.com/frobnitzem/dwork} (GPL v3)
\item {\tt mpi-list}: Mpi-list is released in the pypi package index, and contains full documentation
and functionality tests. \href{https://github.com/frobnitzem/mpi\_list}{https://github.com/frobnitzem/mpi\_list} (MIT license)
\item Workflow Timings: \href{https://code.ornl.gov/99R/workflow\_timings}{https://code.ornl.gov/99R/workflow\_timings} (CC4-BY-SA)
\end{itemize}
All cases contain example use instructions, while {\tt mpi-list} and {\tt dwork} have more extensive tests and documentation.  Usage code is provided in Fig. 1 for {\tt pmake}, Fig. 2 for {\tt dwork} and Fig. 3 for {\tt mpi-list}.  Full scripts to reproduce the results presented here are present in the workflow timings repository linked above.

\section*{Acknowledgments}

This research was sponsored in part by the Laboratory Directed Research and Development Program at Oak Ridge National Laboratory (ORNL), which is managed by UT-Battelle, LLC, for the U.S. Department of Energy (DOE) under Contract No. DE-AC05-00OR22725. This work also used resources, services, and support provided via the COVID-19 HPC Consortium
(https://covid19-hpc-consortium.org/), which is a unique private-public effort to bring together government, industry, and academic leaders who are volunteering free compute time and resources in support of COVID-19 research, and used
resources of the Oak Ridge Leadership Computing Facility at the Oak Ridge National Laboratory, which is supported by the Office of Science of the U.S. Department of Energy under Contract No. DE-AC05-00OR22725.

\end{document}